# A topological phase buried in a chalcogenide superlattice monitored by a helicity dependent Kerr measurement

*Richarj Mondal,*[,1] *Yuki Aihara,*[1] *Yuta Saito,*[2] *Paul Fons,*[2] *Alexander V. Kolobov,*[2] *Junji Tominaga*[2] *& Muneaki Hase*[,1,2]

[1]Division of Applied Physics, Faculty of Pure and Applied Sciences, University of Tsukuba, 1-1-1 Tennodai, Tsukuba 305-8573, Japan.

[2]Nanoelectronics Research Institute, National Institute of Advanced Industrial Science and Technology, Tsukuba Central 5, 1-1-1 Higashi, Tsukuba 305-8565, Japan.

## ABSTRACT

**Chalcogenide superlattices (SL), formed by the alternate stacking of GeTe and $Sb_2Te_3$ layers, also referred to as interfacial phase change memory (iPCM), are a leading candidate for spin based memory device applications. Theoretically, the iPCM structure it has been predicted to form a 3D topological insulator or Dirac semimetal depending on the constituent layer thicknesses. Here, we experimentally investigate the topological insulating nature of chalcogenide SLs using a helicity dependent time-resolved Kerr measurement. The helicity dependent Kerr signal is observed to exhibit a four cycle oscillation with $\pi/2$ periodicity suggesting the existence of a Dirac-like cone in some chalcogenide SLs. Furthermore, we found that increasing the thickness of the GeTe layer dramatically changes the periodicity, indicating a phase transition from a Dirac semimetal into a trivial insulator. Our results demonstrate that thickness-tuned chalcogenide SLs can play an important role in the manipulation of topological states, which may open up new possibilities for spintronic devices based on chalcogenide SLs.**

**1. Introduction**

Topological insulators (TIs),[1-5] the newest class of quantum materials in material science, are promising candidates for spintronic devices,[3] quantum computation,[6] and optical devices[7] as the spins of the TI surface states (SSs) are polarized and orthogonally locked to the crystal momentum.[8,9] TI states are characterized by metallic SSs which are protected from backscattering by time reversal symmetry (TRS),[8-10] while the finite energy band gap of the bulk behaves like an insulator. $Bi_2Se_3$, $Sb_2Te_3$, and $Bi_2Te_3$, well known as TI materials, were discovered over the past several years.[2] Discovering and exploring new TI or related topological materials has been a hotspot in materials research as part of the drive to utilize them in a variety of device applications. However, the alternate stacking of a TI with a normal insulator (NI) gives rise to an intriguing phenomena associated with topological phases in solids. Burkov *et al.* have theoretically predicted that the alternate stacking of a TI with a NI will form a topological superlattice (SL), which can be either TI, NI or a Weyl semimetal depending on how the topological surface modes are coupled by tunneling.[11] Recently, TI based SLs, such as $PbSe/Bi_2Se_3$[12,13] and $GeTe/Sb_2Te_3$,[14-16] have been experimentally examined and such SLs constitute a promising platform to control diverse topological phases.

$GeTe/Sb_2Te_3$, a TI based chalcogenide SL, which consists of the alternate stacking of the ferroelectric NI GeTe with the TI $Sb_2Te_3$, is a leading candidate for the next generation non-volatile electronic memory[17] due to its ultra-low switching energy and outstanding durability of over $10^9$ cycles as compared with the $10^6$ cycles of conventional $Ge_2Sb_2Te_5$ (GST225) based phase change random access memory (PCRAM).[15] This newly introduced SL structure, also referred to as interfacial phase change memory (iPCM),[15,16] a leading candidate for spin based memory device applications,[16] is crystalline in both the RESET (low-conductivity) and SET (high conductivity) phases. This chalcogenide SL has been investigated experimentally and was found to exhibit interesting physical properties, which are thought to be related to the topological nature of the RESET phase.[16,18,19] Theoretical predictions based on *ab initio* calculations have shown that GeTe/Sb$_2$Te$_3$ SLs become either TIs, Dirac semimetals, or NIs depending upon the thickness of the individual blocks of GeTe and Sb$_2$Te$_3$.[20-22] Inversion of the electronic bands occurs due to hybridization across the GeTe/Sb$_2$Te$_3$ interfaces, and the strength of the hybridization across the interface can be controlled by varying the layer thickness resulting in a transition between a Dirac semimetal, NI states or a Wyle semimetal.[12,20,22] However, to date, there has been no direct experimental evidence for a Dirac-like electronic energy band structure corresponding to spin polarized linearly dispersed SSs forming a Dirac cone[5,23] in chalcogenide SLs. The detection of a Dirac-like cone in chalcogenide SL is, however, challenging because the Dirac-like cone is hidden by the heterostructure. In general, topological SSs have been observed using angle-resolved photoemission spectroscopy (ARPES),[3,4,24] but this surface sensitive technique is only applicable to the SS located in the uppermost surface, making it difficult to detect the Dirac-like cone interfacial states in the bulk. Also another difficulty is a tendency of *p*-type conduction in GeTe/Sb$_2$Te$_3$ SLs, which hinders the direct observation of the Dirac-like cone as it is located above the Fermi energy. Recently, a novel technique, helicity dependent time-resolved Kerr spectroscopy, has been used to detect the presence of the Dirac-like cone based on the optical Kerr effect (OKE) in the prototypical TIs, Sb$_2$Te$_3$ and Bi$_2$Te$_3$.[25] Some advantages of this technique include, a simple experimental setup in which measurements can be performed in a transmission or reflection geometry in air and at room temperature without ultra high vacuum, contact free optical measurements, and high sensitivity to the Dirac-like cone structures on the surface or hidden from conventional techniques in the bulk.

In this work, we unravel the buried topological interface states of GeTe/Sb$_2$Te$_3$ chalcogenide SLs. We have used helicity dependent time resolved Kerr spectroscopy to detect the Dirac-like cone. The helicity dependent Kerr signal shows a four cycle oscillation with a

π/2 periodicity, which can be attributed to an OKE dominated by an interfacial Dirac-like cone. Furthermore, we have investigated variations in the topological phases by varying the GeTe layer thickness and found indications of a phase transition from a Dirac semimetal into a NI state with increasing GeTe layer thickness.

## 2. Experimental Section

**2.1. *Sample Fabrication:*** The samples used in the present study were thin films of $[(GeTe)_n/(Sb_2Te_3)_1]_3$ SL (where *n* is an integer), which were comprised of three periods of alternately grown GeTe and $Sb_2Te_3$ layers. The samples were grown on Si (100) substrates using helicon wave radio-frequency magnetron sputtering at 230 °C.[26,27] As-deposited (RESET state) chalcogenide SL samples with varying GeTe layer thickness over the range of 0.72 nm to 8.6 nm were grown while the $Sb_2Te_3$ layer thickness was fixed at 1.0 nm. A conventional GST225 polycrystalline (*fcc*-phase) film with a thickness of 40 nm was used for comparison. All samples were capped by a 20 nm thick layer of $ZnS$-$SiO_2$ to prevent oxidation. The ZnS-$SiO_2$ layer is transparent to near infrared light and therefore makes no contribution to the observed signal.

**2.2. *Time-resolved optical Kerr measurement:*** Optical measurements were carried out using a reflection geometry pump-probe setup (see Figure 1) which consisted of a Ti:sapphire laser system that generated a train of < 20 fs pulses at a wavelength of 830 nm (≈ 1.5 eV) at an 80 MHz repetition rate. The optical penetration depth at 830 nm was calculated to be ~ 20 nm,[28] which is greater than the sample thickness up to the moderate GeTe layer thickness sample (4.3 nm). Thus the effects of the optical penetration depth do not play an important role due to the presence of homogeneous optical excitation over the entire sample thickness, except for the thickest sample (GeTe layer thickness of 8.6 nm). The laser beam was split into two parts, hereafter referred to as the pump and the probe, with average powers of 140 mW and 2 mW, respectively. The pump and the probe pulses were co-focused onto the sample to a spot size 70 $\mu$m diameter. The probe beam was incident at the angle of ~10º with respect to the normal of the sample surface and the pump beam was incident closely to the normal as shown in Figure 1. The probe pulse was *p*-polarized, while the polarization of the pump pulse was varied continuously from linearly to circularly polarized and back to a linearly polarized state by rotating a quarter-wave-plate (QWP). The time delay between the pump and probe pulses was introduced by a fast scan delay operated at a frequency of 19.5 Hz.[25, 29] The pump induced change in the Kerr rotation of the probe pulses was monitored using balanced silicon photo diodes as a function of the pump-probe delay. The differential Kerr rotational signal was fed to a low-noise current amplifier to amplify the signal. The amplified signal was integrated by a digital oscilloscope over 700 scans to improve the signal-to-noise ratio. All measurements were performed in air at room temperature.

## 3. Results and Discussion

The transient Kerr rotation signals ($\Delta\theta_k$) as a function of time delay observed for different pump polarization states, i.e., various QWP angles, are shown in Figure 2 for chalcogenide SL and conventional GST225 polycrystalline samples. Near zero time delay, an instantaneous change in the $\Delta\theta_k$ signal appears due to the photo-excitation of spin polarized electrons induced by the pump. The amplitude of the $\Delta\theta_k$ signal changes sign with the helicity of the pump polarization, and can be interpreted in terms of the inverse Faraday effect (IFE).[30] For the chalcogenide SL, the peak amplitude of the $\Delta\theta_k$ signal as a function of the QWP angle is surprisingly different from that of GST225 as can be seen in Figure 2. To get more insight into the origin of this difference, the peak amplitude of the $\Delta\theta_k$ signal is plotted as a function of the pump polarization state as shown in Figure 3. The $\Delta\theta_k$ signal shows a periodic oscillation with the polarization of the pump for both samples. Interestingly, the chalcogenide SL sample exhibited four oscillation periods with a periodicity of $\pi/2$, while two oscillation periods with a periodicity of $\pi$ were observed for the GST225 sample. The polarization dependent $\Delta\theta_k$ signal can be qualitatively expressed as a combination of specular IFE and specular OKE contributions; a detailed discussion can be found in ref. [25],

$$\Delta\theta_k(\alpha) = A(L\sin 4\alpha + C\sin 2\alpha) + D, \qquad (1)$$

$$\text{where } L = Re\left(\frac{\chi^{(2)}\cdot\chi^{(2)}}{2n(1-n)^2}\right), \qquad (2)$$

$$C = -Im\left(\frac{\chi^{(3)}}{2n(1-n)^2}\right). \qquad (3)$$

Here, $\alpha$ is the angle of the QWP, $A$ is a constant term related to the pump intensity, $D$ is a polarization independent background, and $n$ is the refractive index of the medium. The first term in brackets, $L\sin 4\alpha$, gives rise to an oscillation with a periodicity of $\pi/2$, which is attributed to the second order cascading ($\chi^{(2)}\cdot\chi^{(2)}$) OKE, where the second-order susceptibility $\chi^{(2)}$ associated with the topological band structure induced by strong spin-orbit (SO) interaction. The second term $C\sin 2\alpha$, which gives rise to oscillations with a period of $\pi$, is associated with the IFE, due to the third-order susceptibility $\chi^{(3)}$. The experimental observations can be well fit using Eq. (1), as shown by the solid line in Figure 3.

For the case of the chalcogenide SL, it was found that $L \gg C$, implying that the specular OKE dominates the IFE, resulting in a symmetric four cycle oscillation. The OKE contributes to $\Delta\theta_k$ in the presence of strong SO coupling, which leads to the formation of a Dirac-like cone similar to the Dirac-cone in TI formed due to SO coupling.[1,31,32] On the other hand, it is assumed that spatial inversion symmetry at the interfaces is broken by strain effects,[33] leading a nonzero $\chi^{(2)}$, while TRS is preserved. Therefore $\chi^{(2)}$ at the interfaces will contribute to the OKE signal via a cascading nonlinear process.[25] In the present study, $[(GeTe)_n/(Sb_2Te_3)_1]_3$ SL consists of a fixed 1 nm thick of $Sb_2Te_3$ layer, a normal insulator,[25,34] in combination with a varying thickness ferroelectric GeTe insulator. It was confirmed that both crystalline $Sb_2Te_3$ and GeTe samples show a signal with a periodicity of $\pi$ up to a certain thickness.[25] Nevertheless, we have observed a strong oscillation with a period of $\pi/2$ for a thin SL film, confirming the fact that the $\Delta\theta_k$ signal is not simply the sum of contributions from the GeTe and $Sb_2Te_3$ layers, but comes from the interfaces of the hybridized SL structure where Dirac like-cones may exist.[16,19–21] Thus, a four period oscillation having a π/2 periodicity as a function of the light helicity suggests the existence of the Dirac-like cone or surface Rashba states[35] at interfaces due to the strong SO coupling present in hybridized chalcogenide SLs. Theoretically it has been predicted that a SL formed by [(BiTeI)$_l$(Bi$_2$Te$_3$)$_m$] blocks, where $l$ and $m$ are integers, can act as a TI due to the presence of a strong Rashba effect.[36] This is in contrast to conventional GST225 alloy, where the IFE dominates resulting in larger $C$ than $L$ values leading to a two period oscillation with π periodicity. This π periodic oscillation with helicity is also observed in conventional insulators, such as amorphous GeTe [25] and the semiconductor GaAs.[37] Note that, here we have reported on a randomly oriented 40 nm thick polycrystalline GST225 resulting in oscillations with a period of π, although a single crystalline GST225 is expected to be a topological insulator as reported.[38,39] Our observations on the GeTe/Sb$_2$Te$_3$ SL are also consistent with the results reported by Makino *at el*, who identified topological-like surface and interface states in GeTe/Sb$_2$Te$_3$ SL using THz absorption spectroscopy.[40] Furthermore, theoretical studies based on *ab initio* computer simulations have predicted that the RESET phase of GeTe/Sb$_2$Te$_3$ SL has a Dirac-like cone on the surface as well as at internal interfaces.[20-22]

Figure 4(a) shows the helicity dependence of the peak $\Delta\theta_k$ amplitude for various GeTe layer thicknesses. The helicity dependent $\Delta\theta_k$ exhibits an oscillation with a π/2 period with increasing GeTe thickness. Interestingly, with increasing GeTe layers thickness, the symmetric periodic oscillations change to asymmetric periodic oscillations and in addition the amplitude of the periodic oscillations decrease. The helicity dependent periodic oscillations can be well fit by Eq. (1) as shown by the solid line in Figure 4 (a) for different GeTe layer thicknesses. The fitting parameter $L$, attributed to the OKE contribution is associated with the presence of topological band structure, and $C$ contribution associated with the IFE, have been extracted using Eq. (1). The extracted $L$ and $C$ values as a function of the GeTe layer thickness are shown in Figure 4(b). For very thin GeTe layer thicknesses, the $L$ (OKE) value is very large and dominates the $C$ term (IFE) (here the $C$ value is almost negligible) giving rise to a symmetric oscillation with a period of $\pi/2$ that suggests the existence of a Dirac-like cone.[25] With increasing GeTe layer thickness, the OKE contribution to $\Delta\theta_k$ decreases while the contribution of the IFE component is enhanced, resulting in a decrease in the $L$ value and a concomitant asymptotic increase in the $C$ value, respectively, suggesting that the Dirac-like cone approaching the band gap opening at the $\Gamma$ point. Therefore, a clear indication of a phase transition from a Dirac semimetal toward a NI is observed with increasing GeTe layer thickness. For the case of the thick GeTe layer (~ 8.6 nm), the total thickness of a GeTe/Sb$_2$Te$_3$ SL sample is $\approx$ 28.8 nm, a value larger than the optical penetration depth of $\approx$ 20 nm. The $\Delta\theta_k$ signal change in Fig. 4(a) is, however, not simply due to a reduction in number of detectable layers of the SL sample, since a drastic decrease was already observed when the GeTe thickness was increased from 0.72 to 4.3 nm, during which the $L$ value changed by $\geq$ 50%. In this case the total thickness of the layers was within the optical penetration depth, i.e. the same number of SL layers was monitored. Furthermore, the effect of the bottom surface on the $\Delta\theta_k$ signal is negligibly small due to the use of a reflection geometry.

Theoretically it has been suggested that four different layer stacking orders in GeTe/Sb$_2$Te$_3$ SLs can exist.[20] As a result of phase equilibrium in thermodynamics, the RESET phase can have the stacking orders [··· Te-Sb-Te-Ge-Te-Ge-Te-Sb-Te ···] (Kooi; K)[41] or [Ge-Te ··· Te-Sb-Te-Sb-Te ··· Te-Ge] (inverted Petrov; iP),[42] while the SET phase have the stacking orders [··· Te-Ge-Te-Sb-Te-Sb-Te-Ge-Te ···] (Petrov; P) or [Te-Sb-Te-Sb-Te ··· Te-Ge-Te-Ge] (Ferro; F).[20] DFT simulations suggest both the iP and F phases have a Dirac-like cone, while the P and K phases do not have Dirac-like cones, although for F phase it does not appear at the $\Gamma$ point as for the iP phase, but it appears along the line between A and H.[43] The RESET phase of an as-deposited chalcogenide SL is thus a mixture of K and iP phases. In this case, it is expected that the iP phase would show a larger *L* value, while the *C* term will dominate for the K phase. Here, we have seen that the *L* value decreases with increasing GeTe layer thickness. A possible reason for this behavior is hybridization across the interface of GeTe/$Sb_2Te_3$. An electronic band inversion occurs across the GeTe/$Sb_2Te_3$ interfaces due to hybridization effects and the strength of the hybridization effects decreases with increasing GeTe layer thickness resulting in the gradual disappearance of the Dirac like cones. The resultant band structure for the case of the thick GeTe layer (~ 8.6 nm) is dominated by bulk Rashba-type GeTe bands without a Dirac point.[44,45] On the other hand, the $\pi/2$-periodicity (*L* value) term from the GeTe layer is expected to have contribution dominantly from the surface Rashba bands as in Ref.[44]. If our observation was just simply the sum of signals from the GeTe and $Sb_2Te_3$ layers and if the surface Rashba effect dominates the value of *L*, an increase in the $\pi/2$-periodicity (*L* value) with increasing GeTe layer thickness should be seen. This is, however, not the case here. This fact implies that the signal from the GeTe/$Sb_2Te_3$ samples is not simply the sum of contributions from the GeTe and $Sb_2Te_3$ layers, but from the hybridized system, in which Dirac cones can exist at the interfaces.[16,19–21] The nonzero *L* value for the thickest sample (GeTe layer thickness of 8.6 nm) may include contributions from the bulk Rashba-type bands in GeTe layer and thus the two contributions may be distinguished from Fig. 4. The increase in the value of the *C* term can thus be explained by the increase in the bulk Rashba-type GeTe bands. However, the real structure of the iPCM may be more complicated than the model GeTe/$Sb_2Te_3$ SL structure as intermixing of the layers has been reported.[46] Even though intermixing takes place, since the resultant ternary compounds still possess topological insulating properties,[47] it is expected that the large *L* value can be attributed to the topological properties of the GeTe/$Sb_2Te_3$ SL, although more experimental and theoretical works will be required to fully understand the topological properties.

## 4. Conclusion

To conclude, we have investigated topological phases in chalcogenide superlattices comprised of alternately stacked layers of GeTe and $Sb_2Te_3$, also known as iPCM, using the helicity dependence of time-resolved Kerr measurements. A Kerr rotational signal with the helicity of the pump photons exhibits four oscillations with $\pi/2$ periodicity and is associated with a Dirac-like cone, suggesting chalcogenide SLs behave like a Dirac semimetal. Furthermore, we have investigated the GeTe layer thickness dependence of the Kerr signal and found that a phase transition from a Dirac semimetal to a trivial insulating phase occurs with increasing GeTe layer thickness. Such a phase transition can be understood based on the extent of hybridization across the GeTe/$Sb_2Te_3$ interface. Our results may open up new pathways for developing spintronic devices based on iPCM.

## AUTHOR INFORMATION

**Corresponding Authors**

*Email: rmondal@bk.tsukuba.ac.jp (R.M.).
*Email: mhase@bk.tsukuba.ac.jp (M.H.)

**Author Contributions**

M. H. organized this project. R. M. and Y. A. performed experiments. R. M. analyzed the data. Y. S. and J. T. fabricated the sample. R. M., Y. S., P. F., A. V. K., J. T. and M. H. discussed the results. R.M. and M.H. co-wrote the manuscript.

**Funding**

This research was financially supported by JST-CREST (No. JPMJCR14F1), Japan and JSPS-KAKENHI (No. 17H02908), Japan.

## ACKNOWLEDGMENT

We acknowledge Ms. R. Kondou for sample preparation.

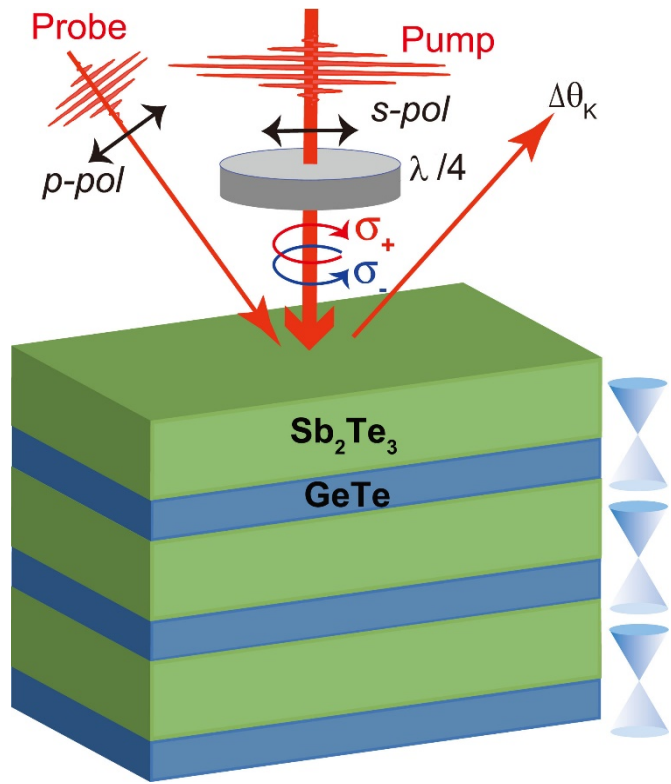

**Figure 1**. Schematic design of a GeTe/Sb$_2$Te$_3$ chalcogenide SL investigated by time resolved Kerr rotational spectroscopy.

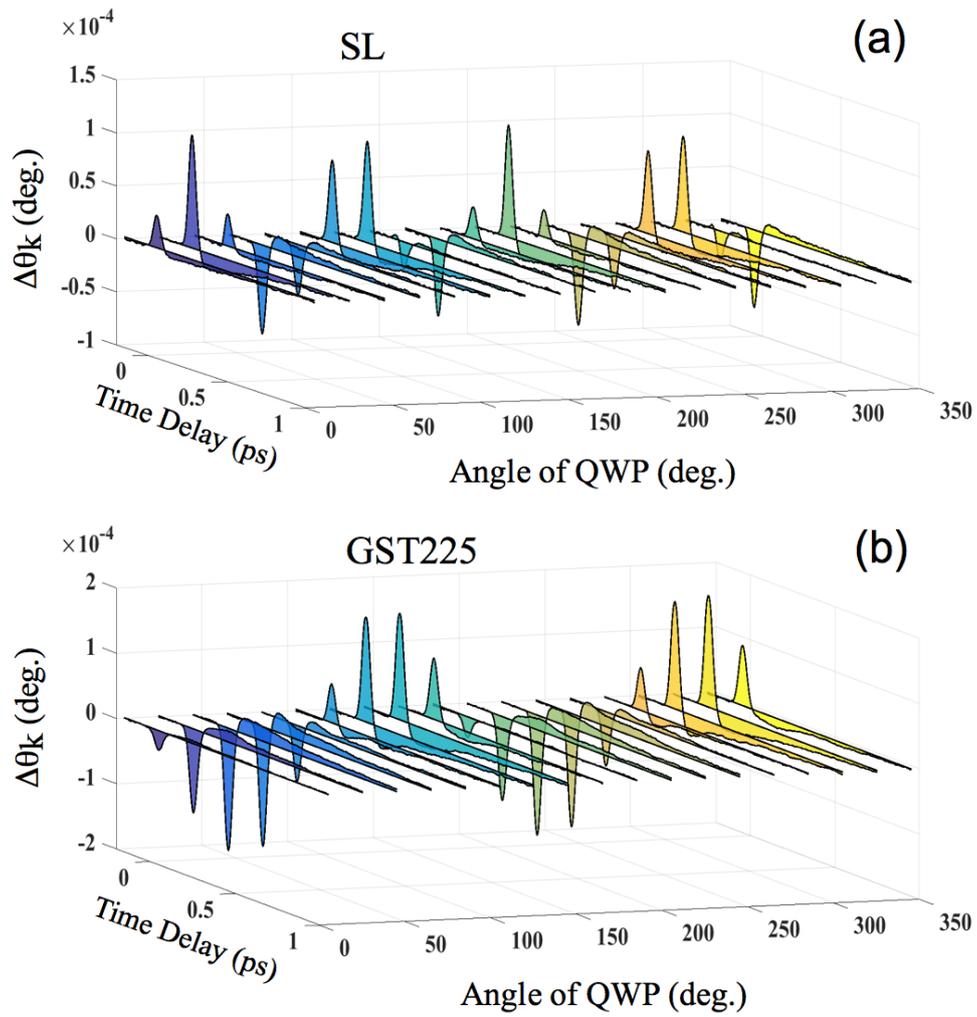

**Figure 2**. Transient Kerr rotational signal ($\Delta\theta_k$) with various angles of quarter wave plate (QWP) measured for (a) GeTe/Sb$_2$Te$_3$ chalcogenide SL and (b) conventional GST225 films.

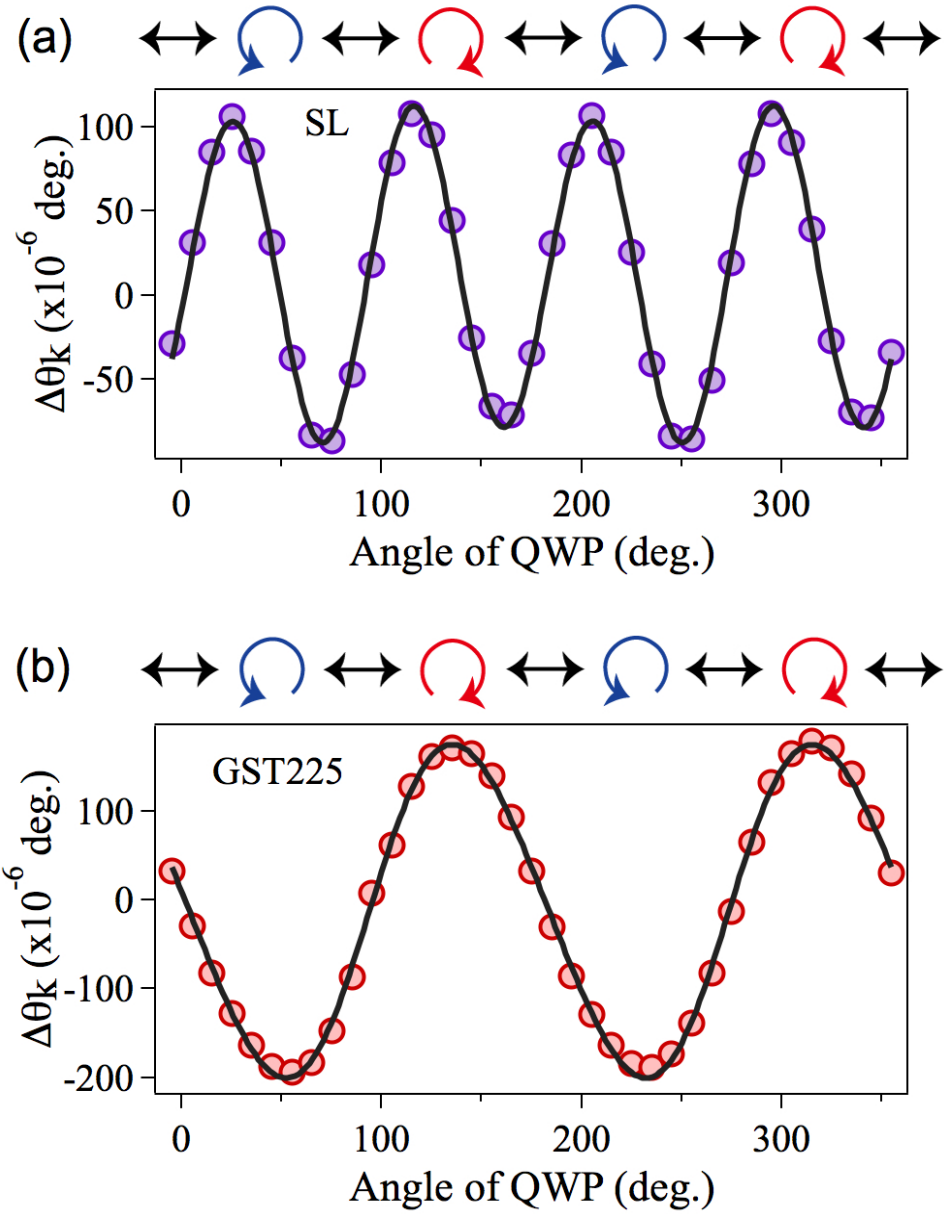

**Figure 3**. Pump polarization dependent peak amplitude of $\Delta\theta_k$ signal for (a) GeTe/Sb$_2$Te$_3$ chalcogenide SL and (b) GST225 films. The solid line represents a fit to the data using Eq. (1).

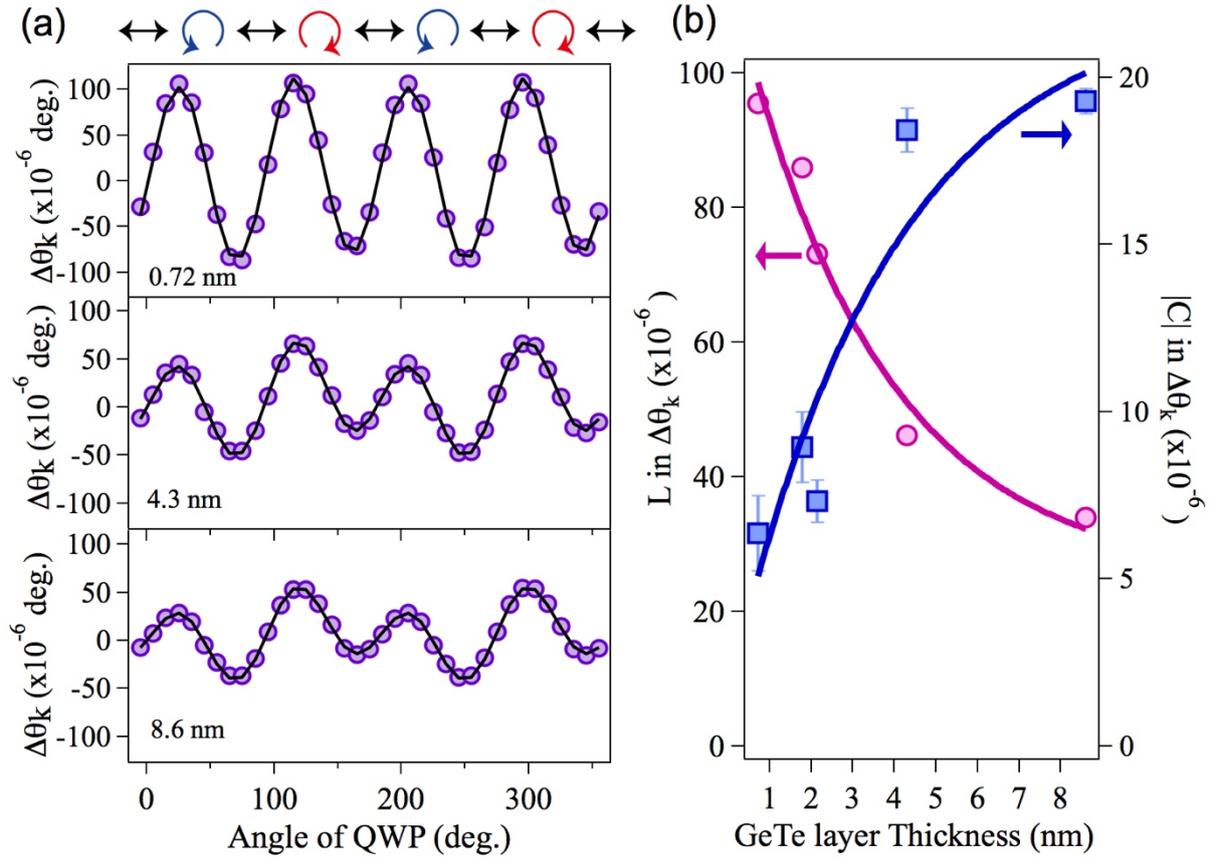

**Figure 4**. (a) The $\Delta\theta_k$ signal as a function of the pump helicity for three different thicknesses of the GeTe sublayer corresponding to 0.72, 4.3 and 8.6 nm, respectively. (b) The *L* and *C* values obtained using Eq. (1) are plotted as a function of the GeTe layer thickness. The solid line is a guide to the eyes.